# ON THE DEFINITION OF THE MEASUREMENT UNIT FOR EXTREME QUANTITY VALUES: SOME CONSIDERATIONS ON THE CASE OF TEMPERATURE AND THE KELVIN SCALE

Franco Pavese [1]

[1] *Torino, Italy*

*E-mail (corresponding author): frpavese@gmail.com*

**Abstract**

Many quantities are attributed a range of values that can apparently extend to infinity (on one side or both sides). In this respect, the definitions of their measurement units do not place any constraint to the maximum (or minimum) value for their validity. In general, that happens because those extreme values are far from being reached on the earth, or presently in experiments. However, since the same units are used also in fields of physics, chemistry or technology where they could occur—namely in the description of the universe in one sense, and in pico-nanoscale or particle physics in another sense—the issue of extreme values (not in statistical meaning here) is not irrelevant. The question placed and discussed in this paper is whether the present kelvin scale, based on Lord Kelvin's second definition (our currently accepted concept of temperature), applies over a full range between bounds (0, ∞) or not, and about the concept of temperature in itself in the extremes regions. The aim, however, is not to provide an answer, but to suggest there are difficulties with the application of current concepts at extremes of temperature.

**Keywords:** kelvin scale; Lord Kelvin proposals; extreme temperatures; temperature definition; unit definition

# 1 Introduction

Many quantities have a range of values that can apparently extend to infinity (on one side or both sides). In this respect, the definition of their measurement units does not place any constraint to the maximum (or minimum) value for its validity. In general, that happens because those extreme values are far from being reached on the Earth or presently in experiments. However, since the same units are used also in fields of physics, chemistry or technology where they could occur—namely in the description of the universe in one sense, and in nanoscale or particle physics in another sense—the issue of extreme values (not in statistical meaning here) is not irrelevant.

Physicists have already raised the point long since. For example, almost one century ago the physicist Bridgman discussed the issue for time ("What is the meaning in saying that an electron when colliding with a certain atom is brought to rest in $10^{-18}$ s?" in 1927) [1] and length ("What is possible meaning of the statement that the diameter of an electron is $10^{-13}$ cm?" in 1955) [2], then opting for operational definitions.

The same issue also attracted the philosophers of science. For example, in a recent book [3], Chang asked "Is a definition valid for a quantity full range?", and introduced the concept of "metrological extension". He proposed: "In order for the extension to be valid, there are two different conditions to be satisfied:

–*Conformity*. If the concept possesses any pre-existing meaning in the new domain, the new standard should conform to that meaning;

–*Overlap*. If the original standards and the new standard have an overlapping domain of application, they should yield measurement results that are consistent with each other"

(also called "compatibility requirement").

These conditions may not be achieved, and "patching up disconnected standards" could occur instead. However, the apparently best solution would be when "one could seek a single standard to cover the entire range": however, one has to admit that for "material standards … no matter how broadly a standard is applicable, it will have its limits".

Thus, in principle, only a theory-based definition—i.e. a model-based definition with operational method(s) ("realisations") available—might satisfy the necessary conditions.

The above applies also to *temperature* for both its asymptotes for extreme low and extreme high values.

Figure 1 depicts the temperature range with the fields of physics and their applications indicated, and the kelvin and Celsius scales superimposed (present experimental limits: ≈$10^{-9}$ K and ≈$3 \cdot 10^9$ K).

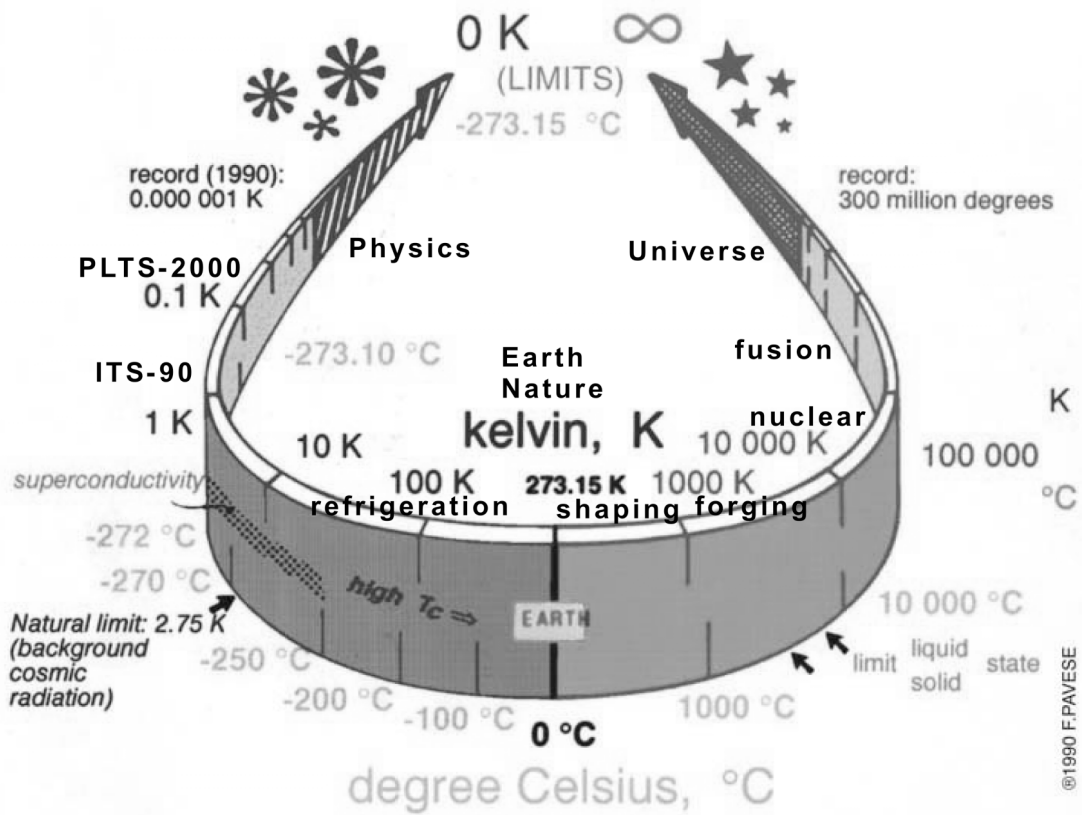

Fig.1. Temperature ranges (modified from [4])

The question placed in this paper is whether the *present* kelvin scale, based on Lord Kelvin's second definition, applies over the full (indefinite) range or not. The aim of this paper is to make a historical recall and tackle some of the modern views, but not to provide an answer, suggesting instead doubts about whether current concepts of temperature may apply at temperature extremes.

The issue of the upper extreme, +∞, will not be further developed in the following. It is enough to inform here that, quite far from experimental science, contemporary models of physical cosmology postulate an "absolute hot", i.e. that the highest possible temperature is the Planck temperature, 1.416 808(33) ×$10^{32}$ K [5] (energy of the Planck mass for $k_B = 1$). Above about $10^{32}$ K, particle energies become so large that gravitational forces between them would become as strong as other fundamental forces according to current theories. There is not yet a scientific theory for the behaviour of matter at these energies. A quantum theory of gravity would be required.[1] [7]

## 2 Historical bases: The nature and rational of Lord Kelvin's proposals

2.1 First Definition

The basis of the first Kelvin definition in 1848 was the work of Carnot [8] about the maximum efficiency of mechanical work that can be obtained from a thermal cycle using steam. [9] He was attracted by Carnot's speculation that his cycle (see Fig. 2) was valid irrespective to the substance used, so becoming 'universal'. In fact, his first definition can be written as:

$$W = (T'_2 - T'_1) Q \quad \text{(for Carnot's function } \mu(T) = const) \tag{1}$$

where $W$ = mechanical work, $Q$ = spent heat.

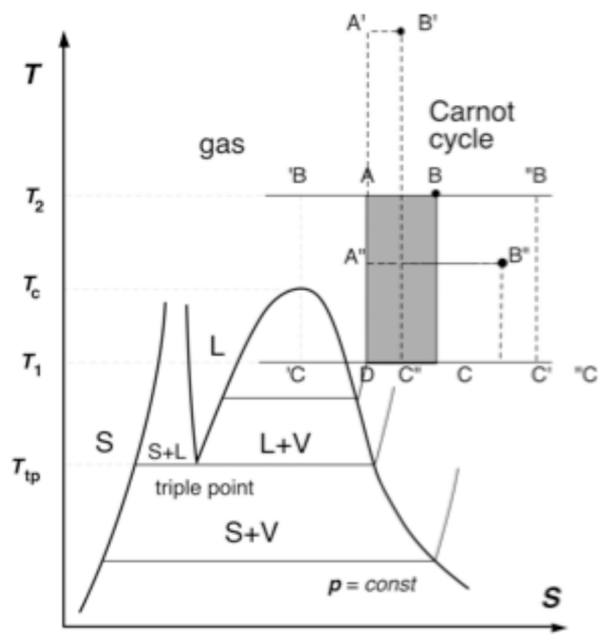

Fig.2. Carnot cycle for a real gas. (from [4])

Thus the definition is tied to (mechanical) work, historically linked to the use of a gaseous media and heat for producing work, i.e., "power motion".

---

[1] "The point at which our physical theories run into most serious difficulties is that where matter reaches a temperature of approximately $10^{32}$ degrees, also known as Planck's temperature. The extreme density of radiation emitted at this temperature creates a disproportionately intense field of gravity. To go even farther back, a quantum theory of gravity would be necessary, but such a theory has yet to be written." [6]

2.2 Second Definition

Subsequently, Lord Kelvin was attracted by the studies of Joule about the conservation of energy (1840), and jointly issued in 1854 a second definition [10]:

$$T^{II}_1 / T^{II}_0 = Q_1 / Q_0. \tag{2}$$

It fits the ideal gas model (real gases can fit the assumption in a restricted temperature range), a limited operability range because gas phase only exists between ≈ 10 K (only solids below, apart helium) and a few thousands degrees (plasma above). Until 1926, the first official application was based on a hydrogen gas thermometer.

The $T^{II}$ definition is a constant-energy one: "one kelvin is equal to the change of thermodynamic temperature [$T^{II}$] that results in a change of thermal energy $kT$ by 1.380 648 52 ×10$^{-23}$ J". [11]

The process bringing Lord Kelvin to his second definition is described in details in [12], namely after the bases for the first and second law of thermodynamics where established by Joules and Clausius, and without the consequence being specifically noted that it involves a zero for the absolute temperature. The relationship between the two Kelvin's definitions is an intriguing one [13]:

$$T^{I} = J \log T^{II} + const \tag{3}$$

where $J$ is Joules' mechanical equivalent of heat ($W \Leftrightarrow Q$ constant = 4.1868 J cal$^{-1}$). When Lord Kelvin's papers were reprinted in 1882, he added a note in which he admitted that his $T^{I}$ defines a logarithmic-type scale.

Table 1. Temperature values in Kelvin's first definition (L) and second definition (K).

| $T$/K | $T$/L | $T$/K | $T$/L |
|---|---|---|---|
| → ∞ [a] | → ∞ [a] | 100 | −322 |
| 1 000 000 | 2 630 | 10 | −1 060 |
| 100 000 | 1 892 | 1 | −1 798 |
| 10 000 | 1 154 | 0.1 | −2 536 |
| 1 000 | 416 | 0.01 | −3 274 |
| 373.15 | 100 | 0.001 | −4 012 |
| 273.15 | 0 | → 0 [a] | → −∞ [a] |

[a] But see text about the possibility to approach the asymptote.

Thus, in principle, $T^{\,\mathrm{I}}$ would match the whole range in a uniform way, where 273.16 $\mathrm{K^{II}} \rightarrow 0\ \mathrm{K^{I}}$ (for a Celsius-related choice of the defining constants), and with an equilibrated emphasis for lower temperatures *approaching* $-\infty\ \mathrm{K^{II}}$ instead of 0 $\mathrm{K^{I}}$. Contrarily to $T^{\,\mathrm{II}}$, $T^{\,\mathrm{I}}$ *is not linear in energy*. Table 1 shows some temperature values in both $T^{\,\mathrm{II}}$ and $T^{\,\mathrm{I}}$ scales: the fact that the values in $T^{\,\mathrm{I}}$ are different from those in $T^{\,\mathrm{II}}$ and all previously used scales in the most common range of everyday temperature measurements is probably a historical reason why $T^{\,\mathrm{I}}$ was never adopted.

2.3 Work versus Energy

One can ask if mechanical work does really fit the needs of thermometry.

Actually, since the Caloric concept was defeated, the modern way-out is to use rather Energy, inclusive of Mechanical Work and Heat.

Energy is a "subtle concept" (from the title of an interesting book on this subject [14]), possibly too much subtle and pervasive, and not so easy to define. For replacing "living force", Young proposed the term "energy" since 1807. Then Lord Kelvin introduced it formally only in 1851 (a bit later, in 1865, Clausius introduced the term 'entropy'). Feynman popular definition is: "Energy is that-which-is-conserved". For Coopersmith [14], "The energy of a system is the capacity of the system to do work"—a definition that requires the concept of 'force', not anymore popular in some branches of physics ("the concept of force is conspicuously absent from our [physical] most advanced formulations of the basic laws", see, e.g., [15]).

Again for Coopersmith [14], "energy has extensive (entropy) and intensive (temperature) attributes", not necessarily correct for temperature.

The proposed New SI [11] definition endorses an energy-based temperature, where $T^{\,\mathrm{II}}$ is linear in energy, as indicated before: $\Delta T = 1$ K for $\Delta Q = k_{\mathrm{B}}$, with $[k_{\mathrm{B}}] = [\mathrm{J\ K^{-1}}]$, the unit of a quantity called heat capacity.

3 Some views on extreme values on the $\boldsymbol{T^{\,\mathrm{II}}}$ kelvin scale

For the high extreme see the end of the Introduction.

3.1 Low extreme

Figure 3 shows a well-known picture of statistical mechanics, on which the $\boldsymbol{T^{\,\mathrm{II}}}$ kelvin scale is presently based, at the lowest temperatures where the statistics splits into different possibilities; note that there classical Boltzmann statistics are just limiting values of either of the Fermi and Bose branches at low occupancy of available quantum states.. In the figure, arrows are added toward 0 K, indicating an asymptotic approach.

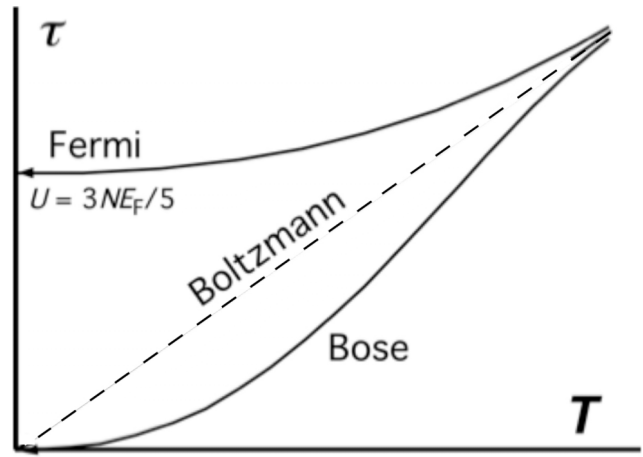

Fig. 3. Statistical models toward 0 K. (modified from [4])

This fact was commented since many years. For example Simon in 1955 [16], while appreciating the Lord Kelvin *First Definition*,[2] noted specific basic problems related to the Second Definition in approaching the "absolute zero":

- The modern justification of the choice of $T^{II}$ is the kinetic theory and statistical mechanics. However, for $T^{II} \to 0$ one reaches a point where the statistical hypotheses are not anymore respected (and before that fluctuations occur);
- The above theories are normally dealing with material's lattice, while for $T^{II} \to 0$ one needs to distinguish between specific sub-systems.

Therefore, the arrows in Fig. 3 do not actually reach exactly $T^{II} = 0$, but must stop somewhat before. Actually, going toward $T^{II} \to 0$, energy is going toward zero logarithmically—and one may wonder if any state can reach the level $E =: 0$—apart possibly the universe whole energy balance, according to some theories (e.g., [17]).

**4 Two contemporary views on possible temperature concepts in future**

Here two examples of these studies are reported.

---

[2] Incidentally, in the recent literature one can find other examples of illustrations, and possible redeeming, of the Lord Kelvin first definition [40–44].

## 4.1 Temperature in nanothermodynamics and the quantum frame

The recent extension of experimental work and technologies to very small dimensions (nanotechnologies) and to very low temperatures (nano-temperatures), prompted new problems and the need of rethinking the very concept of temperature in the lowest range.

As already anticipated, below some dimension scales the concept of temperature does not apply to the whole system. In those cases the concept of "local temperature" is introduced, and one should study the "minimal length scales for the existence of local temperature" (see, e.g. [18–20]).

In these studies it is claimed, for example, that "This length scale is found to be constant for temperatures above the Debye temperature and proportional to $T^{-3}$ below" so that "high temperatures can exist quite locally, while low temperatures exist on larger scales only" and, e.g., "in quasi 1-dimensional systems, like carbon-nanotubes, room temperatures (300 K) exist on length scales of 1 μm, while very low temperatures (10 K) can only exist on scales larger than 1 mm" [18].

More generally about nanothermodynamics, the term was introduced by T.L. Hill in 2000 [21], who already wrote on the issue in 1963-64 concerning "small systems" [22]. The issue was initially prompted by studies on thermodynamic "fluctuations", experimentally observed only since 1992, before the nanotechnology field started to develop.

Several theories have been developed [20, 23], like the non-extensive statistical mechanics, Hill's theory, the tensorial approach.

These studies also involve an effort to reconcile the quantum with the classical thermodynamics [24], with controversial positions (see, e.g., [25, 26]). In [25] similarity between quantum mechanics and thermodynamics is discussed. It is found that if the Clausius equality is imposed on the Shannon entropy and the analogue of the quantity of heat, then the value of the Shannon entropy comes to formally coincide with that of the von Neumann entropy of the canonical density matrix, and pure-state quantum mechanics apparently transmutes into quantum thermodynamics. The corresponding quantum Carnot cycle of a simple two-state model of a particle confined in a one-dimensional infinite potential well was studied. By imposing the Clausius equality the pure-state, quantum mechanics transmutes into genuine quantum thermodynamics, if $k_B$ and $T$ are regarded as the Boltzmann constant and real temperature, respectively. Its efficiency is shown to be identical to the classical one.

However, the authors in [26] contended that the statement is incorrect. In particular, they proved that the state at the beginning of the cycle is *mixed* due to the process of measuring energy. The imposition of the Clausius equality allows the connection between quantum mechanics and thermodynamics, thus resulting in quantum thermodynamics.

An experimental evidence of connection of the classical to quantum world has been described in [27].

4.2 Gibbs versus Boltzmann thermodynamic temperature ($T^{II}$) and negative temperatures

This frame is involving the concept of temperature under another perspective, especially prompted by Costantino Tsallis [28], Peter Hängii [30–31] and others, involving the introduction of a Gibbs thermodynamic temperature $T_G$, alternative to the Boltzmann one, an issue still controversial too (e.g., see [32–34]). According to [30] the relationship between the Boltzmann temperature $T_B$ and the Gibbs one $T_G$ is:

$$T_B = T_G/(1-k_B/C), \text{ where } C = (\partial T_G/\partial E)^{-1} \qquad (3)$$

For classical systems with many degrees of freedom, the difference in the value of the temperature based on $S_B$ and $S_G$ is negligible.

Yet, concerning entropy, Table 2 shows the findings in [30], where Gibbs entropy satisfies the three laws, while Boltzmann does not (for the meaning of the symbols in the column $S(E)$ see [30]. Figure 4 shows the meaning of $\omega$ and $\Omega$).

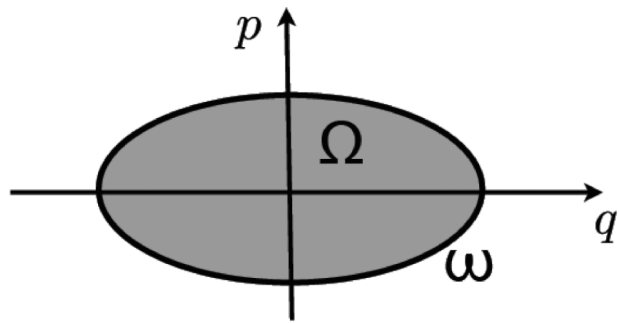

Fig. 4. Standard meaning of $\omega$ and $\Omega$. From [30].

Table 2. Comparison of different definitions of entropy in respect to their accordance to three laws of thermodynamics ($S_B(E) = \ln(\varepsilon\omega)$. $T_B = \omega/\nu >$ or $< 0$, $S_G(E) = \ln(\Omega)$, $T_G = \Omega/\omega > 0$). From [30].

| Entropy | $S(E)$ | Second Law | First Law | Zeroth Law |
| --- | --- | --- | --- | --- |
| Gibbs | $\ln(\Omega)$ | yes | yes | yes |
| Penrose | $\ln(\Omega) + \ln(\Omega_\infty - \Omega) - \ln\Omega_\infty$ | yes | yes | no |
| Complementary Gibbs | $\ln(\Omega_\infty - \Omega)$ | yes | yes | no |
| Differential Boltzmann | $\ln[\Omega(E + \varepsilon) - \Omega(E)]$ | yes | no | no |
| Boltzmann | $\ln(\varepsilon\omega)$ | no | no | no |

However, the authors in [35] argue that Gibbs' entropy fails to satisfy a basic requirement of thermodynamics, namely that when two bodies are in thermal equilibrium they should be at the same temperature, while Boltzmann's one does.

The above discussion involves, as a consequence the acceptance, or not, of negative temperatures (in $T^{II}$)—see, e.g., [36][3]. A recent paper [37], introducing a generalised entropy, intended to be inclusive of Gibbs, Boltzmann and Shannon definitions, supports the latter position

## 4 Lack of a conclusion but not of further scientific studies

This paper did not intend to be a full review of all positions, or to reach firm conclusions, except that in author's opinion there is already evidence that none of the definitions of the official temperature scales is convenient for covering the extreme regions of the range of the intended quantity. The fact that the present experimental studies for temperature are presently to within a $T^{II}$ range of less than ($10^{-10}$ to $10^{10}$) K does not mean that one can be confident that for a long time that there will be no need to worry about the contents of this paper.

---

[3] "Because negative temperature systems can absorb entropy while releasing energy, they give rise to several counterintuitive effects, such as Carnot engines with an efficiency greater than unity. Through a stability analysis for thermodynamic equilibrium, negative temperature states of motional degrees of freedom necessarily possess negative pressure and are thus of fundamental interest to the description of dark energy in cosmology, where negative pressure is required to account for the accelerating expansion of the universe". [36]

For sure one can already state that no "ultimate" solution exist. Any new Kuhn's "revolution" can provide, in one year or in $10^x$ years, new knowledge that, while extending the range where the concept of temperature can be managed, could also innovate, at least partially, in the ranges where today we think to be confident that no innovation is needed.

After all, the previous concepts of *time* (subjective—of the observer), *space* (e.g., "the space of astronomy is not a physical space of meter sticks, but is a space of light waves" [1]) *force* [e.g., 11], *vacuum* [e.g., 38] and even '*ether*' [e.g., 39]—and related ones—have already been revisited.

Accordingly, also relentless studies on the concept of temperature and temperature scales have no reasons for being considered as having become obsolete. The author hopes to have, at least, stimulated the curiosity of young scholars and scientists for new challenges. In this respect a last reference seems appropriate [45].